\documentclass[aps,prl,twocolumn,superscriptaddress,amsmath,amssymb]{revtex4}
\usepackage{graphicx} 
\usepackage{epstopdf} 
\usepackage{dcolumn} 
\usepackage{xcolor} 
\usepackage{amsmath,amssymb} 
\usepackage{hyperref} 
\usepackage[utf8]{inputenc} 
\usepackage[english]{babel} 
\usepackage[displaymath, mathlines]{lineno} 
\usepackage{blindtext} 
\usepackage{ifthen} 
\usepackage{orcidlink} 

\usepackage{setspace}
\usepackage{array}
\graphicspath{{figures/}} 

\newboolean{articletitles}
\setboolean{articletitles}{true} 

\def\belletwo{\mbox{Belle~II}\xspace}
\newcommand{\tev}{\ensuremath{\mathrm{\,Te\kern -0.1em V}}\xspace}
\newcommand{\gev}{\ensuremath{\mathrm{\,Ge\kern -0.1em V}}\xspace}
\newcommand{\mev}{\ensuremath{\mathrm{\,Me\kern -0.1em V}}\xspace}
\newcommand{\kev}{\ensuremath{\mathrm{\,ke\kern -0.1em V}}\xspace}
\newcommand{\ev}{\ensuremath{\mathrm{\,e\kern -0.1em V}}\xspace}
\newcommand{\gevc}{\ensuremath{{\mathrm{\,Ge\kern -0.1em V\!/}c}}\xspace}
\newcommand{\mevc}{\ensuremath{{\mathrm{\,Me\kern -0.1em V\!/}c}}\xspace}
\newcommand{\gevcc}{\ensuremath{{\mathrm{\,Ge\kern -0.1em V\!/}c^2}}\xspace}
\newcommand{\gevgevcccc}{\ensuremath{{\mathrm{\,Ge\kern -0.1em V^2\!/}c^4}}\xspace}
\newcommand{\mevcc}{\ensuremath{{\mathrm{\,Me\kern -0.1em V\!/}c^2}}\xspace}

\def\invfb   {\ensuremath{\mbox{\,fb}^{-1}}\xspace}

\def\Pe      {\ensuremath{e}\xspace}  
\def\epem       {\ensuremath{\Pe^+\Pe^-}\xspace}

\newcommand{\taualpha}{\ensuremath{\tau^-\!\to\!\ell^-\alpha}\xspace}
\newcommand{\taualphae}{\ensuremath{\tau^-\!\to\! e^-\alpha}\xspace}
\newcommand{\taualphamu}{\ensuremath{\tau^-\!\to\!\mu^-\alpha}\xspace}
\newcommand{\tauSM}{\ensuremath{\tau^-\!\to\! \ell^-\bar\nu_\ell \nu_\tau}\xspace}
\newcommand{\taue}{\ensuremath{\tau^-\!\to\! e^-\bar\nu_e \nu_\tau}\xspace}
\newcommand{\taumu}{\ensuremath{\tau^-\!\to\! \mu^-\bar\nu_\mu \nu_\tau}\xspace}
\newcommand{\taualphaSM}{\ensuremath{{\mathcal B}(\tau^-\!\to\! \ell^- \alpha) / {\mathcal B}(\tau^- \!\to\! \ell^- \bar\nu_\ell \nu_\tau)}\xspace}
\newcommand{\taualphaeSM}{\ensuremath{{\mathcal B}(\tau^-\!\to\! e^- \alpha) / {\mathcal B}(\tau^- \!\to\! e^- \bar\nu_e \nu_\tau)}\xspace}
\newcommand{\taualphamuSM}{\ensuremath{{\mathcal B}(\tau^-\!\to\! \mu^- \alpha) / {\mathcal B}(\tau^- \!\to\! \mu^- \bar\nu_\mu \nu_\tau)}\xspace}
\newcommand{\BBe}{\ensuremath{{\mathcal B}_{e \alpha} / {\mathcal B}_{e \bar\nu \nu}}\xspace}
\newcommand{\BBmu}{\ensuremath{{\mathcal B}_{\mu \alpha} / {\mathcal B}_{\mu \bar\nu \nu}}\xspace}
\newcommand{\BBell}{\ensuremath{{\mathcal B}_{\ell \alpha} / {\mathcal B}_{\ell \bar\nu \nu}}\xspace}
\newcommand{\resulte}{\ensuremath{(1.1-9.7) \times 10^{-3}}\xspace}
\newcommand{\resultmu}{\ensuremath{(0.7-12.2) \times 10^{-3}}\xspace}

\newboolean{wordcount}
\setboolean{wordcount}{false} 

\ifthenelse{\boolean{wordcount}}%
{\usepackage{bibentry} 
 \usepackage{comment} 
 \excludecomment{align*} 
 \excludecomment{align} 
 \excludecomment{equation*} 
 \excludecomment{equation} 
 \excludecomment{eqnarray*} 
 \excludecomment{eqnarray} 
 \excludecomment{acknowledgments} 
 \def\maketitle{} 
}{}

\begin{document}

\pacs{}

\title{
\newdimen\origiwtitle
\origiwtitle=\fontdimen2\font
\fontdimen2\font=0.65\origiwtitle
Search for Lepton Flavor Violating $\tau$ Decays to a Lepton and an Invisible Boson at \belletwo
\fontdimen2\font=\origiwtitle}
\ifthenelse{\boolean{wordcount}}{}{
  \author{I.~Adachi\,\orcidlink{0000-0003-2287-0173}} 
  \author{K.~Adamczyk\,\orcidlink{0000-0001-6208-0876}} 
  \author{L.~Aggarwal\,\orcidlink{0000-0002-0909-7537}} 
  \author{H.~Ahmed\,\orcidlink{0000-0003-3976-7498}} 
  \author{H.~Aihara\,\orcidlink{0000-0002-1907-5964}} 
  \author{N.~Akopov\,\orcidlink{0000-0002-4425-2096}} 
  \author{A.~Aloisio\,\orcidlink{0000-0002-3883-6693}} 
  \author{N.~Anh~Ky\,\orcidlink{0000-0003-0471-197X}} 
  \author{D.~M.~Asner\,\orcidlink{0000-0002-1586-5790}} 
  \author{T.~Aushev\,\orcidlink{0000-0002-6347-7055}} 
% ...
\collaboration{The Belle II Collaboration}

}

\begin{abstract}
We search for lepton-flavor-violating \taualphae{} and \taualphamu{} decays, where $\alpha$ is an invisible spin-0 boson. 
The search uses electron-positron collisions at 10.58~\gev center-of-mass energy with an integrated luminosity of 62.8~\invfb, produced by the SuperKEKB collider and collected with the \belletwo{} detector.
We search for an excess in the lepton-energy spectrum of the known \taue{} and \taumu{} decays.
We report 95\% confidence-level upper limits on the branching-fraction ratio \taualphaeSM{} in the range \resulte{} and on \taualphamuSM{} in the range \resultmu{} for $\alpha$ masses between 0 and 1.6~\gevcc. These results provide the most stringent bounds on invisible boson production from $\tau$ decays.
\end{abstract}

\maketitle

The standard model of particle physics successfully describes the electromagnetic, weak, and strong interactions and classifies all known elementary particles. However, phenomena such as neutrino oscillations, long-standing discrepancies between expectations and observations such as the muon magnetic-moment anomaly, and indirect evidence of dark matter clearly indicate that the standard model is incomplete. Many extensions of the standard model that attempt to incorporate these phenomena require new bosons that are candidates for dark matter or that explain the muon's anomalous magnetic moment~\cite{Grinstein:1985rt}.

\newdimen\origiwt
\origiwt=\fontdimen2\font
\fontdimen2\font=0.8\origiwt
Decays of $\tau$ leptons into final states involving light, beyond-the-standard-model bosons that are not directly detectable (invisible) are predicted in models with, e.g., axionlike particles~\cite{Berezhiani:1989fp,Calibbi:2020jvd, Bauer:2019gfk, Cornella:2019uxs}. These bosons are collectively referred to as $\alpha$ in this work. A direct search in \taualpha, where $\ell^-$ indicates $e^-$ or $\mu^-$, can probe theories beyond the standard model with high sensitivity (charge-conjugated decays are implied throughout). \mbox{This process} was previously searched for by the MARK III ~\cite{MARK-III:1985iqj} and ARGUS~\cite{Albrecht:1995ht} collaborations. The current best upper limits on the \taualpha{} branching fractions, relative to the corresponding standard-model leptonic decays, are $\taualphaeSM < (6\textup{--}36) \times 10^{-3}$ and $\taualphamuSM < (3\textup{--}34) \times 10^{-3}$ at the $95 \%$ confidence level (C.L.) where the range indicates their dependence on the $\alpha$ mass in the (0--1.6)$\gevcc{}$~\cite{Albrecht:1995ht} range. 
\fontdimen2\font=\origiwt

We report a search for the lepton-flavor-violating {\taualpha{}} decay involving a spin-0 $\alpha$ boson, 
where both the $\alpha$ and its decay products are invisible. We use data from \epem{} collisions produced in 2019 and 2020 by the SuperKEKB collider at KEK~\cite{Akai:2018mbz}. The data, corresponding to an integrated luminosity of 62.8~\invfb~\cite{Belle-II:2019usr}, were recorded by the \belletwo{} detector~\cite{Abe:2010gxa} at a center-of-mass energy of $\sqrt{s}=10.58$~\gev and contain 57.7 million $e^+e^-\to\tau^+\tau^-$ events.

The Belle II detector consists of several subdetectors arranged in a cylindrical structure around the $e^+e^-$ interaction point. 
The $z$ axis of the laboratory frame is defined as the detector solenoid axis, with the positive direction in the direction of the electron beam. 
The polar angle $\theta$ and the transverse plane are defined according to this axis.
Charged-particle trajectories are reconstructed by a tracking system consisting of a two-layer silicon-pixel detector, surrounded by a four-layer double-sided silicon-strip detector and then a central drift chamber (CDC). Only 15\% of the second pixel layer was installed when the data used in this work were collected. 
Outside the CDC, time-of-propagation and aerogel ring-imaging Cherenkov detectors cover the $31^{\circ} < \theta < 128^{\circ}$ and  $14^{\circ} < \theta < 30^{\circ}$ polar ranges, respectively. 
The electromagnetic calorimeter (ECL) that serves to reconstruct photons and identify electrons fills the remaining volume inside a superconducting solenoid that generates a 1.5 T uniform, axial magnetic field.
A dedicated subdetector to identify $K_L^0$ mesons and muons is located at the outermost radius of the detector. 

In the center-of-mass frame, $\tau$ pairs are produced back to back; thus, decay products of each $\tau$ are isolated from the others and contained in opposite hemispheres. The boundary between those hemispheres is the plane perpendicular to the $\tau$ direction, which can be experimentally approximated by the thrust axis $\hat{t}$, i.e., the vector that maximizes the thrust value $\sum{|\hat{t}\cdot \vec{p}_i^{\rm \, c.m.}|}/\sum{|\vec{p}_i^{\rm \, c.m.}|}$, where $\vec{p}_i^{\rm \, c.m.}$ is the momentum of each final-state particle in the center-of-mass frame~\cite{Brandt:1964sa,Farhi:1977sg}. 
We define the \emph{tag} hemisphere as that containing three charged particles, which originate from $\tau^- \to h^- h^+ h^- \nu_\tau$ ($h=\pi, K$), and require that the other hemisphere, the \emph{signal} hemisphere, contain only one charged particle.

In $\tau$ decays from $\tau$-pair production, the \taualpha{} has an identical visible topology to that of $\tau^- \to \ell^- \bar\nu_\ell \nu_\tau $, the latter thus being an irreducible background. However, in \taualpha{} the magnitude of the lepton momentum depends only on the $\alpha$ mass. The two-body decay provides a distinctive signature for our signal above the spectrum of $\tau^- \to \ell^- \bar\nu_\ell \nu_\tau $, where the lepton momentum has a broad distribution.

We cannot determine the $\tau$ momentum from the observed particles directly; we can, however, approximate its energy in the center-of-mass frame as $\sqrt{s}/2$ (neglecting the initial-state radiation) and its direction as being opposite to the three hadrons on the tag side, $\hat{p}_\tau \approx - \vec{p}_{3h} / |\vec{p}_{3h}|$. From these we construct the $\tau$ \emph{pseudo rest frame}~\cite{Albrecht:1995ht}. 

We search for \taualpha{} by looking for an excess of events above the spectrum of \mbox{$\tau^-\to\ell^-\bar\nu_\ell\nu_\tau $} normalized lepton energy~\cite{Ali:1977de}
\begin{equation}
x_\ell \equiv \frac{E^*_{\ell}}{m_{\tau}c^2/2}\,,
\end{equation}
where $E^*_{\ell}$ is the energy of the charged lepton in the $\tau$ pseudo rest frame. We then measure the branching-fraction ratio $\BBell\equiv\taualphaSM$, using $\tau^- \to \ell^- \bar\nu_\ell \nu_\tau$ as a normalization channel.  

The search uses an online event selection (trigger) that requires events with at least three localized energy deposits in the ECL (clusters). One of these clusters must have an energy above 0.3~\gev and the rest above 0.1~\gev. In addition, we require a topology inconsistent with Bhabha scattering.
The efficiency of this trigger is measured in experimental data with respect to independent triggers based on the number of particles reconstructed in the CDC; simulated distributions are then scaled by this efficiency. The trigger efficiencies are on average $97\%$ and $86\%$ for the electron and muon channels, respectively.

We select $\tau$-pair candidates by requiring the event to contain exactly four charged particles with zero total charge, each displaced from the average interaction space point by less than 3~cm in the $z$-axis direction and less than 1~cm in the transverse plane. The particle in the signal hemisphere must be identified as either an electron or a muon by combining the information from all subdetectors into a global discriminator similar to a likelihood ratio.
Each charged particle in the tag hemisphere must satisfy the condition $E_{\rm ECL}/p \leq 0.8$ to reject electron contamination. Here, $E_{\rm ECL}$ is the particle energy measured in the ECL and $p$ the magnitude of its momentum measured in the tracker, both in the laboratory frame.

Several $e^+e^-$ final states contribute to the spectra as background: $q\bar{q}$ with $q = u, d, s,c$ (hadronic), $\ell^+\ell^-\gamma$ (dileptonic), and $e^+e^-\ell^+\ell^-$,  $e^+e^-h^+h^-$ (two-photon). We use simulated events to determine the criteria to suppress these backgrounds. We use KKMC to simulate $\tau^+\tau^-$, $q\bar{q}$, and $\mu^+\mu^-(\gamma)$ production~\cite{Jadach:1999vf, Jadach:2000ir}; BabaYaga@NLO for $\epem (\gamma)$~\cite{Balossini:2006wc, Balossini:2008xr, CarloniCalame:2003yt, CarloniCalame:2001ny,CarloniCalame:2000pz}; and AAFH and TREPS for nonradiative two-photon production~\cite{BERENDS1985421,BERENDS1985441,BERENDS1986285,Uehara:1996bgt}. Standard-model $\tau$ decays are handled by TAUOLA~\cite{Chrzaszcz:2016fte}  and their radiative corrections by PHOTOS~\cite{Barberio:1990ms}, while \taualpha{} decays are simulated with PYTHIA8.2~\cite{Sjostrand:2014zea} for $\alpha$ mass values of 0.0, 0.5, 0.7, 1.0, 1.2, 1.4, and 1.6~\gevcc and zero $\alpha$ spin. The \belletwo{} analysis software framework~\cite{Kuhr:2018lps,basf2-zenodo} uses the Geant4~\cite{GEANT4:2002zbu} package to simulate the response of the detector. Since we have no prior knowledge of the $\alpha$ mass,
the selection is optimized for the normalization channel using the figure of merit $S / \sqrt{S +B}$, where $S$ is the number of events in the normalization channel, while $B$ is the number of total background events, both taken from simulations.

Backgrounds from $\epem\to q\bar{q}$ are suppressed by rejecting events containing neutral pions and photons. Photons used in $\pi^0$ reconstruction are ECL clusters with energy deposits of at least 0.1~\gev, which must be within the CDC acceptance to ensure they are not matched to any charged particle. 
Displaced clusters from secondary hadronic interactions and multiple clusters deposited by low-momentum charged particles are challenging to model correctly; photons that deposit less than $0.4~\gev$ in the ECL must be at least 40~cm from the nearest charged particle at the inner surface of the ECL to suppress these contributions.
Neutral pions are identified as photon pairs with masses within $[115, 152]~\mevcc$.
Events containing photons satisfying the above conditions, but not used in $\pi^0$ reconstruction and with energy greater than $0.2~\gev$, are also rejected. We reject events from $\epem\to\ell^+\ell^-\gamma$, $e^+e^-\ell^+\ell^-$, and  $e^+e^-h^+h^-$, characterized by low-momentum tag-side charged particles, by sorting the three charged particles in the tag hemisphere by increasing transverse momentum and requiring that they exceed, respectively, 0.08, 0.30, and 0.70~\gevc for \taualphae{} candidates and 0.04, 0.17, and 0.47~\gevc for \taualphamu{} candidates.
We further reject events from $\epem\to\epem\gamma$ and $q\bar{q}$ by restricting the thrust value to ranges consistent with that of $\tau^+\tau^-$ events and from $\epem\to \epem\ell^+\ell^-$ by requiring the event missing momentum, i.e.,\ the negative vector sum of the momenta of all reconstructed particles in the event, to be within a polar angle ($\theta_{\rm miss}$) range where the process can be accurately simulated. In addition we suppress all types of backgrounds by requiring that charged particle trajectories in the tag hemisphere are consistent with a common origin, and that these particles have a mass $M_{3h}$ and center-of-mass-frame energy $E^{\rm c.m.}_{3h}$ consistent with $\tau$ decay kinematics. The ranges for the thrust, $\theta_{\rm miss}$, $E^{\rm c.m.}_{3h}$, and $M_{3h}$ selections are listed in Table~\ref{table:selection}. 
The reconstruction efficiencies in the normalization channels are $12.7\%$ for \taue{} and $16.2\%$ for \taumu{} decays, while the purities are $95.9\%$ and $92.0\%$ respectively. The reconstruction efficiency of \taualphae{} decays depends on the $\alpha$ mass and varies between $9.4\%$ and $13.9\%$. Likewise the efficiency for \taualphamu{} decays varies between $9.1\%$ and $17.4\%$.

\begin{table}[thb]
\footnotesize
\centering
\caption{Requirements on event thrust, missing momentum polar angle, and tag hemisphere particles' total center-of-mass energy and mass.
}
\label{table:selection}
\begin{tabular}{p{0.25\columnwidth}>{\centering\arraybackslash}p{0.35\columnwidth}>{\centering\arraybackslash}p{0.35\columnwidth} }
\hline
\hline
& \taualphae{} & \taualphamu \\
\hline
Thrust & $[0.90, 0.99]$ & $[0.90, 1.00]$\\
$\theta_{\rm miss}$ & $[20^{\circ}, 160^{\circ}]$ & $[20^{\circ}, 160^{\circ}]$\\
$E^{\rm c.m.}_{3h}$ & $[1.2 , 5.3]~\gev$ & $[1.1, 5.3]~\gev$ \\
$M_{3h}$ & $[0.5, 1.7]~\gevcc$ & $[0.4, 1.7]~\gevcc$\\
\hline
\hline
\end{tabular}
\end{table}
 
\begin{figure}[b]
    \centering
    \includegraphics[width=\linewidth]{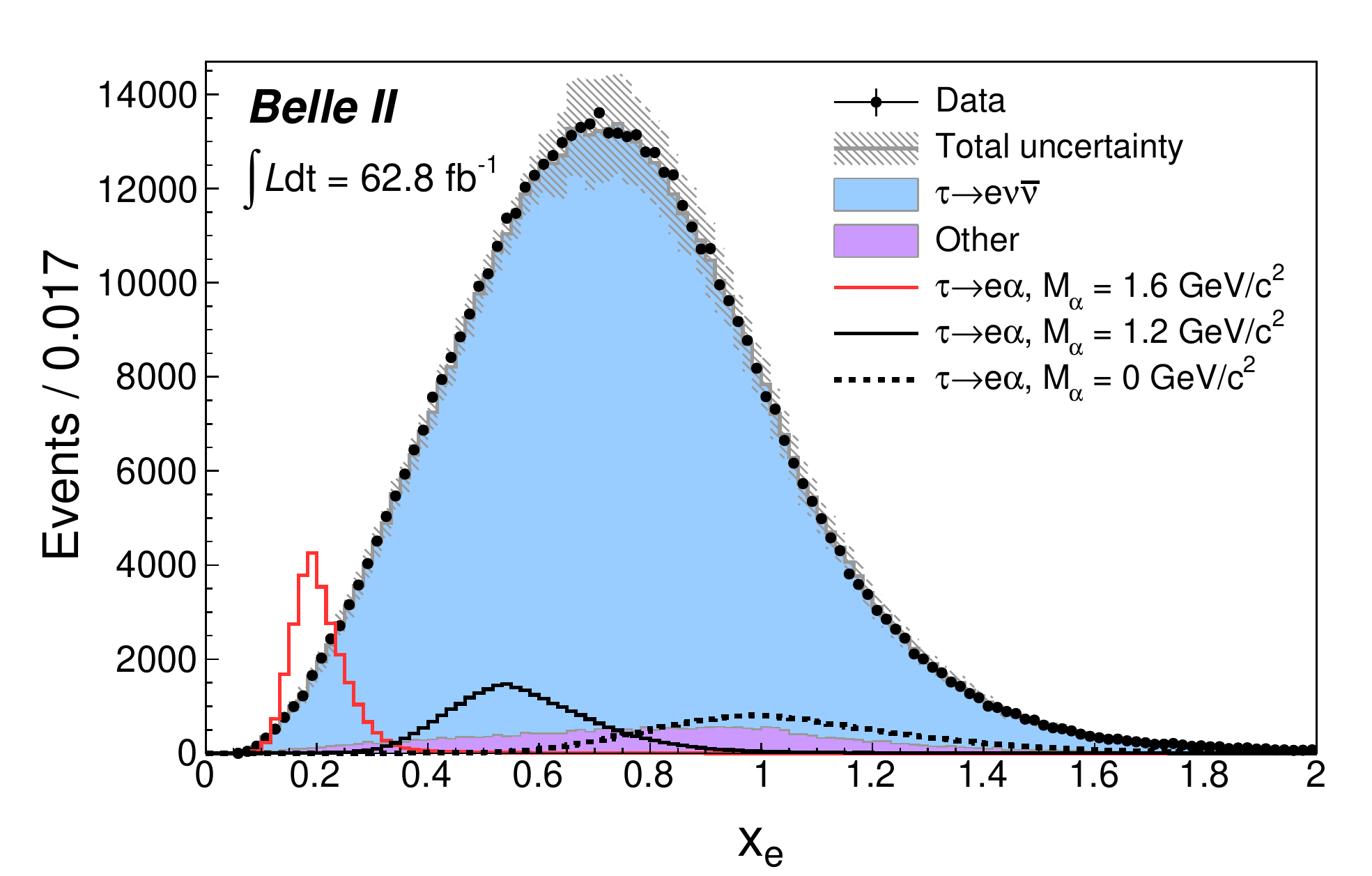}
    \includegraphics[width=\linewidth]{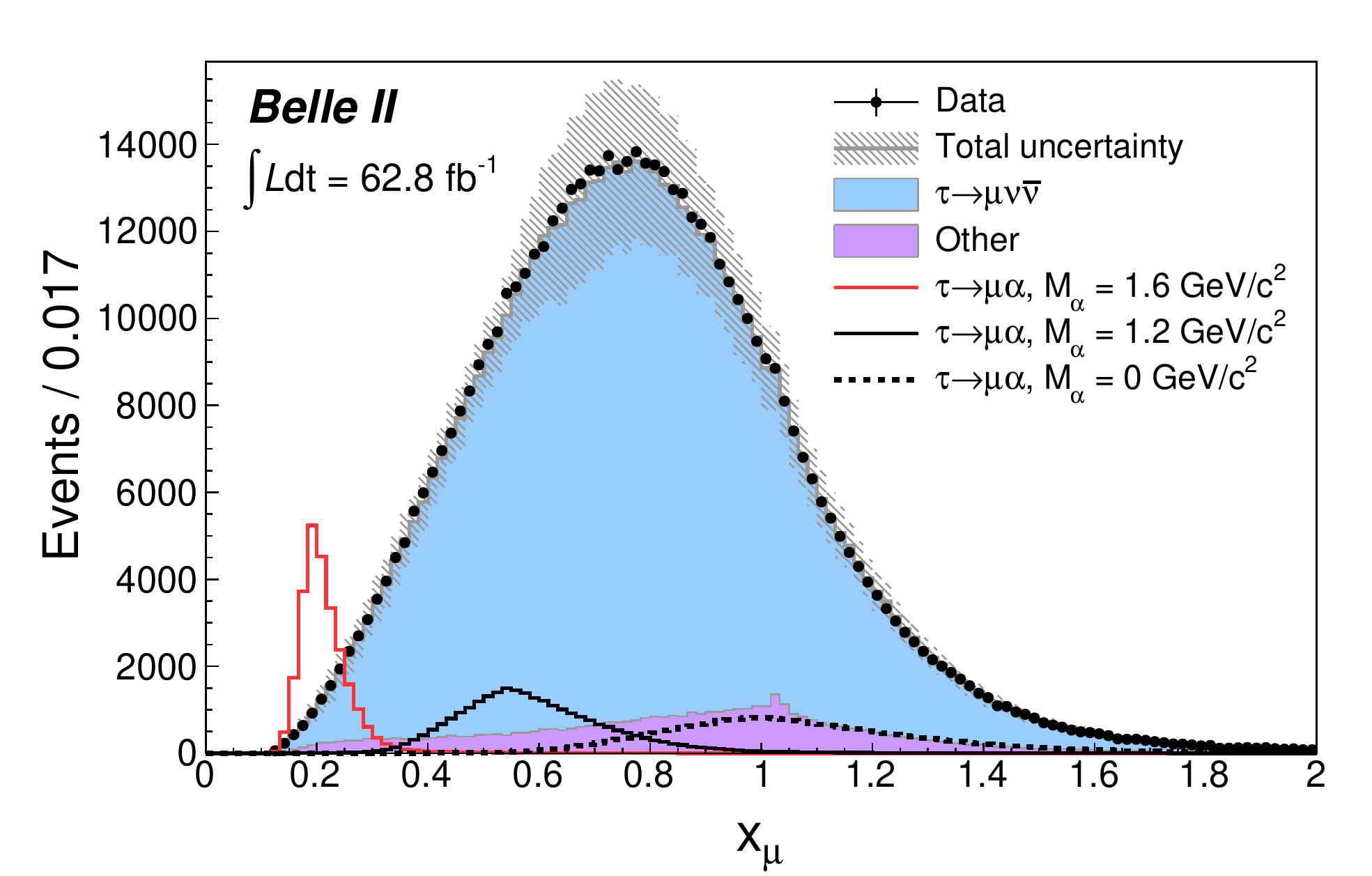}
    \caption{Spectra of $x_\ell$ for electrons (top) and muons (bottom) in simulation and experimental data. Simulated spectra for standard-model processes are shown stacked, 
    with the gray band indicating the total uncertainty, which is dominated by the lepton-identification efficiency uncertainty.
    Remaining background processes other than \tauSM{} contributing to the spectrum are combined together and collectively referred to as ``other''. 
    The distributions for \taualpha{} are shown for three $\alpha$ masses assuming branching-fraction ratios of $5\%$. 
    }
    \label{fig:datamc_x_prf}
\end{figure}

The parameter space defined by the selection criteria is referred to as the signal region. We perform the analysis in a closed-box approach; before examining the $x_\ell$ distribution of experimental data in the signal region, we validate the simulation using variables that are insensitive to the presence of \taualpha{} decay, and study control regions defined by accepting events containing neutral pions or photons, rather than rejecting them. The distributions of $x_\ell$ for events belonging to the signal region are shown in Fig.~\ref{fig:datamc_x_prf}. 

We model each $x_\ell$ spectrum as a sum of contributions from the signal decay, the standard-model leptonic decay, and all other sources of background,

\begin{eqnarray}
\label{eqn:model}
N(x_\ell) &=& N_{\ell\bar\nu\nu}
\,\frac{\epsilon_{\ell\alpha}}{\epsilon_{\ell\nu\nu}} \, \frac{{\mathcal B}_{\ell\alpha}}{{\mathcal B}_{\ell\bar\nu\nu}} \, f_{\ell\alpha}(x_\ell)  \nonumber \\  
     & + & N_{\ell\bar\nu\nu} \, f_{\ell\bar\nu\nu}(x_\ell) + N_{b}  \, f_{b}(x_\ell) \;,
\end{eqnarray}
where the probability density functions $f_{\ell\alpha}$, $f_{\ell\bar\nu\nu}$, and $f_b$ are binned distributions taken from simulations, $N_{\ell\bar\nu\nu}$ and $N_{b}$ are the observed yields, and $\epsilon_{\ell\alpha}/\epsilon_{\ell\nu\nu}$ is the efficiency of observing \taualpha{} decays relative to that for observing $\tau^- \to \ell^- \bar\nu_\ell \nu_\tau $. 

We use RooStats~\cite{Moneta:2010pm} and HistFactory~\cite{Cranmer:2012sba} to fit our model to binned data using extended maximum likelihoods that are functions of the branching-fraction ratio \BBell{} and of $N_{\ell\bar\nu\nu}$ and $N_{b}$.

The leading systematic uncertainties originate from the corrections to the lepton-identification efficiency and particle misidentification rate, based on comparison of calibration samples in data and simulated events. These corrections depend on the momentum and polar angle; their typical ranges are summarized in Table~\ref{table:lepid}. The resulting uncertainties are asymmetric and strongly depend on $x_\ell$; their ranges and averaged values over the standard-model yields are also reported in the same table. The contribution from lepton-identification efficiency partially cancels in the ratio between signal and normalization channels; while the contribution from particle misidentification rates does not, as it affects only other background sources. 

\begin{table}[thb]
\footnotesize
\centering
\caption{Typical ranges for corrections to the lepton-identification efficiencies and misidentification rates, together with ranges for their respective uncertainties and their average values.}
\label{table:lepid}
\newdimen\origiwspc%
\newdimen\origiwstr%
\origiwspc=\fontdimen2\font
\fontdimen2\font=0.6\origiwspc
\begin{tabular}{p{0.38\columnwidth}>{\centering\arraybackslash}p{0.18\columnwidth}>{\centering\arraybackslash}p{0.18\columnwidth}>{\centering\arraybackslash}p{0.19\columnwidth} }
\hline
\hline
 & Correction & Uncertainty & Average \\
 & range & range(\%) & uncert.(\%)\\
\hline
Electron identification & 0.84--1.06 & 0.9--12.6 & $+5.3, -2.9$ \\
Muon identification & 0.63--1.02 & 1.3--32.8 & $+11.7, -1.6$ \\
Electron misidentification & 0.6--6.0 & 4.3--34.6 & $+17.6,-14.7$ \\
Muon misidentification & 0.3--1.5 & 1.4--37.0 & $+18.0,-18.2$ \\
\hline
\hline
\end{tabular}
\fontdimen2\font=1.0\origiwspc
\end{table}

\begin{table*}[th]
\footnotesize
\centering
    \caption{Central values with their uncertainties, 95\% C.L., and 90\% C.L. upper limits (UL) for the branching-fraction ratios \BBe{} (top) and \BBmu{} (bottom) for various masses of the $\alpha$ boson. Corresponding absolute upper limits for ${\mathcal B}(\taualpha)$, computed using standard-model branching fractions from Ref.~\cite{Workman:2022ynf}, are provided in parentheses for convenience.}
    \label{table:upperlimits}
\begin{tabular}{p{0.20\textwidth}>{\centering\arraybackslash}p{0.25\textwidth}>{\centering\arraybackslash}p{0.25\textwidth}>{\centering\arraybackslash}p{0.25\textwidth} }
\hline
\hline\\[-0.8em]
$M_\alpha [\gevcc]$ 
& \hspace{0.2cm} \BBe{}  $(\times 10^{-3})$                      
& \hspace{0.2cm} UL at $95\%$ CL $(\times 10^{-3})$             
& \hspace{0.2cm} UL at $90\%$ CL $(\times 10^{-3})$\\
\\[-0.8em]
\hline
$0.0$  & $-8.1 \pm 3.9$ &$5.3~(0.94)$ &$4.3~(0.76)$\\
$0.5$  & $-0.9 \pm 4.3$ &$7.8~(1.40)$ &$6.5~(1.15)$\\
$0.7$  & $\;\;\:1.7  \pm 4.0$ &$9.0~(1.61)$ &$7.6~(1.36)$\\
$1.0$  & $\;\;\:1.7  \pm 4.2$ &$9.7~(1.73)$ &$8.2~(1.47)$\\
$1.2$  & $-1.1 \pm 2.6$ &$4.5~(0.80)$ &$3.7~(0.66)$\\
$1.4$  & $-0.3 \pm 1.0$ &$1.8~(0.32)$ &$1.5~(0.26)$\\
$1.6$  & $\;\;\:0.2  \pm 0.5$ &$1.1~(0.19)$ &$0.9~(0.16)$\\[0.5em]
\hline
\hline\\[-0.8em]
$M_\alpha [\gevcc]$ 
& \hspace{0.2cm} \BBmu{} $(\times 10^{-3})$                                     
& \hspace{0.2cm} UL at $95\%$ CL $(\times 10^{-3})$ 
& \hspace{0.2cm} UL at $90\%$ CL $(\times 10^{-3})$ \\
\\[-0.8em]
\hline
$0.0$ & $-9.4\pm 3.7$  & $3.4~~(0.59)$ &$2.7~~(0.47)$\\
$0.5$ & $-3.2\pm 3.9$  & $6.2~~(1.07)$ &$5.1~~(0.88)$\\
$0.7$ & $\;\;\:2.7\pm 3.4$  & $9.0~~(1.56)$ &$7.8~~(1.35)$\\
$1.0$ & $\;\;\:1.7\pm 5.4$  & $12.2~~(2.13)\;\:$ &$10.3~~(1.80)\;\:$\\
$1.2$ & $-0.2\pm 2.4$  & $3.6~~(0.62)$ &$2.9~~(0.51)$\\
$1.4$ & $\;\;\:0.9\pm 0.9$  & $2.5~~(0.44)$ &$2.2~~(0.38)$\\
$1.6$ & $-0.3\pm 0.5$  & $0.7~~(0.13)$ &$0.6~~(0.10)$\\
\hline
\hline
\end{tabular}
\end{table*}

Uncertainties from the trigger and $\pi^0$ reconstruction efficiency corrections are also taken into account. Trigger uncertainties range in $0.1\%$--$4\%$ for the electron channel and in $0.2\%$--$1.5\%$ for the muon channel, depending on $x_\ell$. Neutral pion reconstruction efficiency is evaluated from studies on independent samples to be $0.914\pm0.020$. 
Each of these systematic uncertainties is included in the likelihood as an additional shape-correlated nuisance parameter that is assumed to follow a Gaussian distribution. Other sources of uncertainty from track reconstruction efficiency, beam-energy determination, relative reconstruction efficiency, and momentum-scale correction have negligible impact on the results.

Inspection of events in the signal region shows that asymmetrical uncertainties yield unreliable results. We, therefore, revise our definitions and symmetrize their distributions using their greater variation in each bin.

\begin{figure}[htb]
    \centering
    \includegraphics[width=\linewidth]{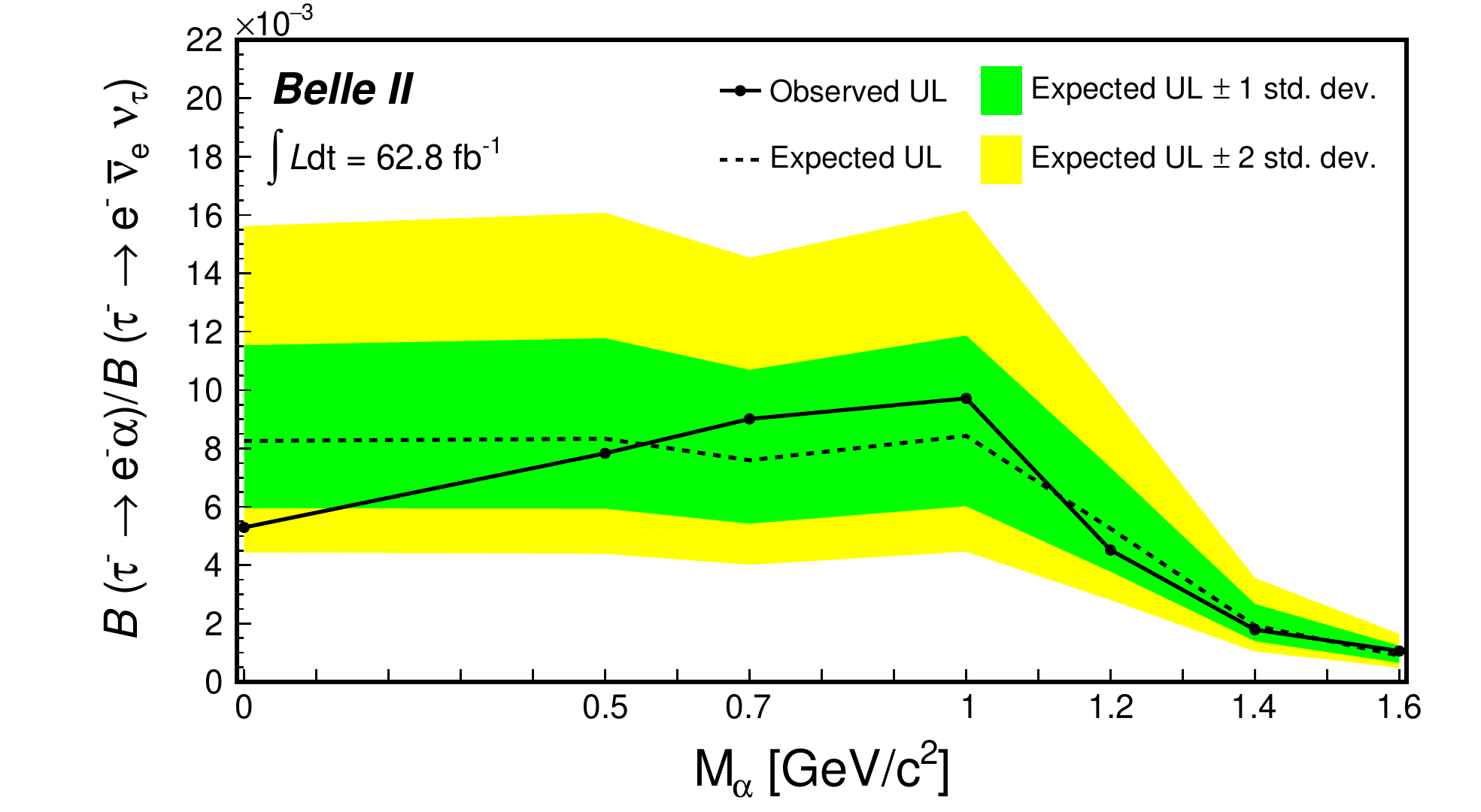}
    \includegraphics[width=\linewidth]{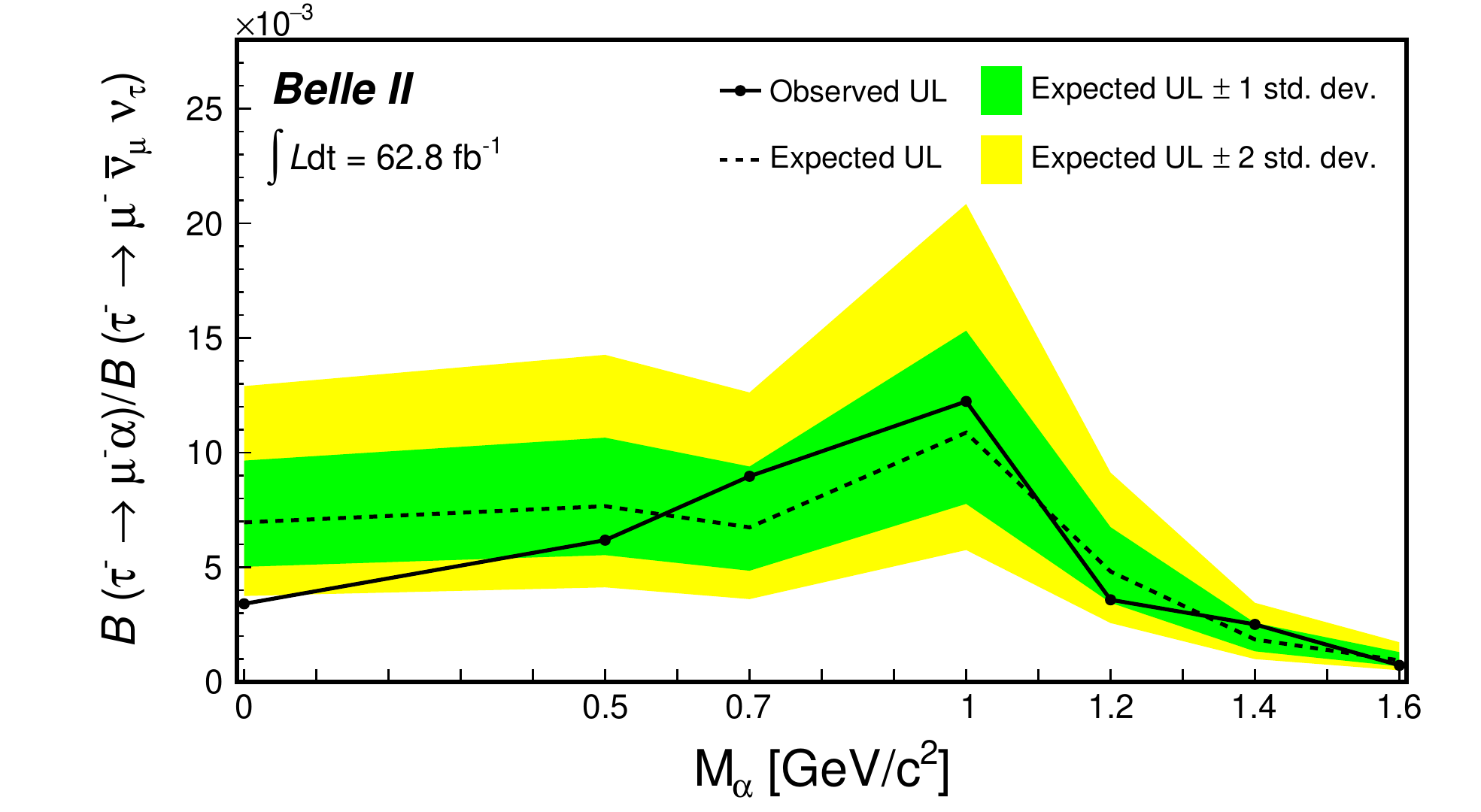}
    \caption{Upper limits at $95\%$ C.L. on the branching-fraction ratios \taualphaeSM{} (top) and \taualphamuSM{} (bottom) as a function of the $\alpha$ mass, as well as their expectations from background-only hypothesis. All values are linearly interpolated between mass points. 
    }
    \label{fig:brazil}
\end{figure}

We observe no significant signal and determine upper limits using the CL$_S$ method~\cite{Read:2002hq}, a modified frequentist approach based on a profile likelihood ratio~\cite{Cowan:2010js}. Figure~\ref{fig:brazil} shows the 95\% C.L. upper limits as well as expectations calculated assuming the background-only hypotheses, ranging in \resulte{} for the electron channel and in \resultmu{} for the muon channel. Systematic uncertainties degrade on average our upper limit sensitivity by approximately $35\%$ in both channels.

The fit results and upper limits are summarized in Table~\ref{table:upperlimits}. The corresponding absolute upper limits for ${\mathcal B}(\taualpha)$, computed using standard-model world-average branching fractions for the reference channel~\cite{Workman:2022ynf}, are also provided for convenience. Our 95\% C.L. limits are 2.2--14 times more stringent than the best previous bounds in~\cite{Albrecht:1995ht}, depending on the value of the $\alpha$ mass.

In conclusion, we search for the lepton-flavor-violating decay \taualpha{} using data collected by the \belletwo{} detector at an \epem{} center-of-mass energy of $10.58$~\gev, corresponding to an integrated luminosity of 62.8~\invfb. We observe no statistically significant signal and set $90\%$ and $95\%$ confidence-level upper limits on the branching-fraction ratios \taualphaSM{}. These constitute the most stringent limits on invisible spin-0 boson production from $\tau$ lepton decays, allowing one to directly constrain standard model extensions (see, e.g., Ref.~\cite{Calibbi:2020jvd}) in ways not otherwise possible outside of collider experiments.

\begin{acknowledgments}
This work, based on data collected using the Belle II detector, which was built and commissioned prior to March 2019, was supported by
Science Committee of the Republic of Armenia Grant No.~20TTCG-1C010;
Australian Research Council and research Grants
No.~DE220100462,
No.~DP180102629,
No.~DP170102389,
No.~DP170102204,
No.~DP150103061,
No.~FT130100303,
No.~FT130100018,
and
No.~FT120100745;
Austrian Federal Ministry of Education, Science and Research,
Austrian Science Fund
No.~P~31361-N36
and
No.~J4625-N,
and
Horizon 2020 ERC Starting Grant No.~947006 ``InterLeptons'';
Natural Sciences and Engineering Research Council of Canada, Compute Canada and CANARIE;
Chinese Academy of Sciences and research Grant No.~QYZDJ-SSW-SLH011,
National Natural Science Foundation of China and research Grants
No.~11521505,
No.~11575017,
No.~11675166,
No.~11761141009,
No.~11705209,
and
No.~11975076,
LiaoNing Revitalization Talents Program under Contract No.~XLYC1807135,
Shanghai Pujiang Program under Grant No.~18PJ1401000,
Shandong Provincial Natural Science Foundation Project~ZR2022JQ02,
and the CAS Center for Excellence in Particle Physics (CCEPP);
the Ministry of Education, Youth, and Sports of the Czech Republic under Contract No.~LTT17020 and
Charles University Grant No.~SVV 260448 and
the Czech Science Foundation Grant No.~22-18469S;
European Research Council, Seventh Framework PIEF-GA-2013-622527,
Horizon 2020 ERC-Advanced Grants No.~267104 and No.~884719,
Horizon 2020 ERC-Consolidator Grant No.~819127,
Horizon 2020 Marie Sklodowska-Curie Grant Agreement No.~700525 "NIOBE"
and
No.~101026516,
and
Horizon 2020 Marie Sklodowska-Curie RISE project JENNIFER2 Grant Agreement No.~822070 (European grants);
L'Institut National de Physique Nucl\'{e}aire et de Physique des Particules (IN2P3) du CNRS (France);
BMBF, DFG, HGF, MPG, and AvH Foundation (Germany);
Department of Atomic Energy under Project Identification No.~RTI 4002 and Department of Science and Technology (India);
Israel Science Foundation Grant No.~2476/17,
U.S.-Israel Binational Science Foundation Grant No.~2016113, and
Israel Ministry of Science Grant No.~3-16543;
Istituto Nazionale di Fisica Nucleare and the research grants BELLE2;
Japan Society for the Promotion of Science, Grant-in-Aid for Scientific Research Grants
No.~16H03968,
No.~16H03993,
No.~16H06492,
No.~16K05323,
No.~17H01133,
No.~17H05405,
No.~18K03621,
No.~18H03710,
No.~18H05226,
No.~19H00682, 
No.~22H00144,
No.~26220706,
and
No.~26400255,
the National Institute of Informatics, and Science Information NETwork 5 (SINET5), 
and
the Ministry of Education, Culture, Sports, Science, and Technology (MEXT) of Japan;  
National Research Foundation (NRF) of Korea Grants
No.~2016R1\-D1A1B\-02012900,
No.~2018R1\-A2B\-3003643,
No.~2018R1\-A6A1A\-06024970,
No.~2018R1\-D1A1B\-07047294,
No.~2019R1\-I1A3A\-01058933,
No.~2022R1\-A2C\-1003993,
and
No.~RS-2022-00197659,
Radiation Science Research Institute,
Foreign Large-size Research Facility Application Supporting project,
the Global Science Experimental Data Hub Center of the Korea Institute of Science and Technology Information
and
KREONET/GLORIAD;
Universiti Malaya RU grant, Akademi Sains Malaysia, and Ministry of Education Malaysia;
Frontiers of Science Program Contracts
No.~FOINS-296,
No.~CB-221329,
No.~CB-236394,
No.~CB-254409,
and
No.~CB-180023, and No.~SEP-CINVESTAV research Grant No.~237 (Mexico);
the Polish Ministry of Science and Higher Education and the National Science Center;
the Ministry of Science and Higher Education of the Russian Federation,
Agreement No.~14.W03.31.0026, and
the HSE University Basic Research Program, Moscow;
University of Tabuk research Grants
No.~S-0256-1438 and No.~S-0280-1439 (Saudi Arabia);
Slovenian Research Agency and research Grants
No.~J1-9124
and
No.~P1-0135;
Agencia Estatal de Investigacion, Spain
Grant No.~RYC2020-029875-I
and
Generalitat Valenciana, Spain
Grant No.~CIDEGENT/2018/020;
Ministry of Science and Technology and research Grants
No.~MOST106-2112-M-002-005-MY3
and
No.~MOST107-2119-M-002-035-MY3,
and the Ministry of Education (Taiwan);
Thailand Center of Excellence in Physics;
TUBITAK ULAKBIM (Turkey);
National Research Foundation of Ukraine, project No.~2020.02/0257,
and
Ministry of Education and Science of Ukraine;
the U.S. National Science Foundation and research Grants
No.~PHY-1913789 
and
No.~PHY-2111604, 
and the U.S. Department of Energy and research Awards
No.~DE-AC06-76RLO1830, 
No.~DE-SC0007983, 
No.~DE-SC0009824, 
No.~DE-SC0009973, 
No.~DE-SC0010007, 
No.~DE-SC0010073, 
No.~DE-SC0010118, 
No.~DE-SC0010504, 
No.~DE-SC0011784, 
No.~DE-SC0012704, 
No.~DE-SC0019230, 
No.~DE-SC0021274, 
No.~DE-SC0022350; 
and
the Vietnam Academy of Science and Technology (VAST) under Grant No.~DL0000.05/21-23.

These acknowledgements are not to be interpreted as an endorsement of any statement made
by any of our institutes, funding agencies, governments, or their representatives.

We thank the SuperKEKB team for delivering high-luminosity collisions;
the KEK cryogenics group for the efficient operation of the detector solenoid magnet;
the KEK computer group and the NII for on-site computing support and SINET6 network support;
and the raw-data centers at BNL, DESY, GridKa, IN2P3, INFN, and the University of Victoria for offsite computing support.

We also thank Pablo Roig Garc\'es for the useful discussion on the interpretation of the results.

\end{acknowledgments}

\ifthenelse{\boolean{wordcount}}%
{ \nobibliography{references} }
{ \bibliography{references} }

\end{document}